\documentstyle[aps,prl,floats,psfig]{revtex}  
\draft  
  
\begin{document}  
  
\twocolumn[\hsize\textwidth\columnwidth\hsize\csname  
@twocolumnfalse\endcsname  
  
\hfill\vbox{  
\hbox{MADPH-01-1232}  
\hbox{AMES-HET-01-09}  
\hbox{hep-ph/0106207}  
\hbox{}}  
  
\title{Unknowns after the SNO Charged-Current Measurement}  
\author{V. Barger$^1$, D. Marfatia$^1$ and K. Whisnant$^2$}  
\address{$^1$Department of Physics, University of Wisconsin--Madison, WI 53706, USA\\  
$^2$Department of Physics and Astronomy, Iowa State University,  
Ames, IA 50011, USA}  
  
\maketitle  
  
\begin{abstract}   
We perform a model-independent analysis of solar neutrino flux rates  
including the recent charged-current measurement at the Sudbury Neutrino  
Observatory (SNO).  We derive a universal sum rule involving SNO and  
SuperKamiokande rates, and show that the SNO neutral-current measurement  
can not fix the fraction of solar $\nu_e$ oscillating to sterile  
neutrinos. The large uncertainty in the SSM $^8$B flux impedes
a determination of the sterile neutrino fraction.  
  
\end{abstract}  
]  
The Solar Neutrino Problem (SNP) is the discrepancy between the   
neutrino flux measured by solar neutrino experiments  
\cite{homestake,sage,gallex,gno,sk,sno} and the predictions of the  
 Standard Solar Model (SSM)~\cite{SSM}. The SNP has defied non-particle  
physics explanations~\cite{pl}. The best-motivated  
solution is massive neutrinos with oscillations of solar    
electron neutrinos to mu and/or tau neutrinos.  
The SuperKamiokande (SK) experiment prefers solutions  
with large mixing between the mass eigenstates~\cite{s}. Very recently, the   
SNO collaboration   
presented initial results of their charged-current (CC) measurement   
from about one year of operation, which again confirms  
the flux-suppression~\cite{sno}. The combination of SNO and SK data   
definitively establishes that the flux-suppression of solar neutrinos is  
of particle physics origin, since it can be inferred that $\nu_{\mu,\tau}$ come  
from the sun~\cite{sno}.  
It is commonly believed that  
measurements of the neutral-current (NC) flux  
in the SNO experiment will decide whether oscillations to  
{\it sterile} neutrinos (that do not possess the SM weak-interaction)   
occur~\cite{smirnov}.  
  
A motivating reason to postulate the existence of sterile neutrinos comes  
from the LSND accelerator experiment~\cite{lsnd} which finds a   
$\nu_\mu \rightarrow \nu_e$ appearance probability of about 0.25\%.   
To explain the solar and atmospheric~\cite{atm} anomalies and the LSND data  
simultaneously, three distinct frequencies of oscillations are required.   
Since with the three known neutrinos   
there are only two independent oscillation frequencies,   
a fourth neutrino must be invoked. However, the invisible width of  
the Z-boson places a constraint on the number of weakly interacting   
neutrinos to be very close to three~\cite{lep}.   
The only way to evade this constraint is   
to require that the fourth neutrino be  
 sterile.   
A recent combined analysis of solar and atmospheric data found that   
the active-sterile  
admixture can take any value between $0$ and $1$ at 99\% C.L. for the  
preferred LMA (Large Mixing Angle) solution to the SNP~\cite{conch}.  
The SNO CC data are inconsistent with maximal mixing to sterile neutrinos   
at the 3.1$\sigma$ level~\cite{sno}. However, SNO did not address arbitrary  
active-sterile admixtures.  
    
In this Letter we perform a neutrino oscillation   
parameter-independent  
analysis of the solar neutrino rates in the  
$^{37}$Cl~\cite{homestake}, $^{71}$Ga~\cite{sage,gallex,gno} and  
SK experiments, and the recent CC measurement at SNO. The $^8$B neutrinos  
represent a large fraction of the neutrinos incident at the SNO, SK  
and $^{37}$Cl experiments, as can be seen from Table~\ref{tab:spectrum}.   
Thus, the $^8$B flux plays a crucial role in the   
interpretation of the results from  
these experiments. Unfortunately, the   
predicted value of the $^8$B flux normalization is quite uncertain  
mainly due to poorly known nuclear cross-sections at low energies~\cite{poor}.   
We find that if the fraction of solar $\nu_e$ that   
oscillate to sterile neutrinos is specified, the data 
determines the normalization of the $^8$B  
solar neutrino flux. Alternatively, if the $^8$B flux normalization   
is assumed to be that of the SSM,   
the range of the sterile neutrino fraction is  
determined. {\it However, the existing solar neutrino rate data and the   
forthcoming SNO NC measurement are not sufficient to determine the   
sterile neutrino content}. We discuss  
the additional measurements that are needed to determine, in a  
model-independent way, the oscillation probabilities and the fraction  
of solar $\nu_e$ that may be oscillating to sterile neutrinos.  
  
\begin{table}[b]  
\caption[]{\label{tab:spectrum}  
Fractional contributions of the high, intermediate and low energy  
neutrinos to the $^{37}$Cl, $^{71}$Ga and SK signals without  
oscillations. The last column gives the $1\sigma$ normalization uncertainty for  
each part of the spectrum.}  
\begin{tabular}{l|l|ccc|c}  
&&&& SK & Norm.\\  
& & $^{37}$Cl & $^{71}$Ga & SNO & Uncertainty\\  
\hline  
High  
& $^8$B, $hep$ & 0.764 & 0.096 & 1.000 & 18.0\%\\  
Inter.  
& $^7$Be, $pep$, $^{15}$O, $^{13}$N & 0.236 & 0.359 & 0.000 & 11.6\%\\  
Low  
& $pp$ & 0.000 & 0.545 & 0.000 & 1.0\%  
\end{tabular}  
\end{table}  
\vskip 0.1in  
\noindent  
\underline{Model-independent analysis.}  
Following the approach of Refs.~\cite{bpw91} and \cite{bmw} (in which  
we made the unique prediction   
$R^{\rm CC}_{\rm SNO}=0.35^{+0.16}_{-0.09}$ for purely active oscillations  
with the SSM $^8$B flux constraint),   
we divide  
the solar neutrino spectrum into three parts: high energy (consisting of  
$^8$B and $hep$ neutrinos), intermediate energy ($^7$Be, $pep$,  
$^{15}$O, and $^{13}$N), and low energy ($pp$). For each class of solar  
neutrino experiment the fractional contribution from each part of the   
unoscillated neutrino spectrum  
 to the expected SSM rate can be calculated  
 (see Table~\ref{tab:spectrum}). We define $P_H$, $P_I$, and  
$P_L$ as the average oscillation probabilities for the high,  
intermediate, and low energy solar neutrinos, respectively. We   
assume that the high-energy solar neutrino flux has absolute  
normalization $\beta_H$ relative to the SSM calculation. If $R$ is the  
measured rate divided by the SSM prediction for a given experiment, then  
with oscillations  
\begin{eqnarray}  
R_{\rm Cl} &=& 0.764 \beta_H P_H + 0.236 P_I \,,  
\label{eq:RCl}\\  
R_{\rm Ga} &=& 0.096 \beta_H P_H + 0.359 P_I + 0.545 P_L \,,  
\label{eq:RGa}\\  
R_{\rm SK} &=& \beta_H P_H + r \beta_H \sin^2\alpha (1 - P_H) \,,  
\label{eq:RSK}\\  
R^{\rm CC}_{\rm SNO} &=& \beta_H P_H \,, 
\label{eq:RSNOCC}  
\end{eqnarray}  
where $r \equiv \sigma_{\nu_\mu, \nu_\tau}/\sigma_{\nu_e}  
\simeq 0.171$ is the ratio of the $\nu_{\mu,\tau}$ to  
$\nu_e$ elastic scattering cross sections on electrons.  
Here $\sin^2\alpha$ is the fraction of  
$\nu_e$ that oscillate to active neutrinos, where $\alpha$ is  
a mixing angle in the four-neutrino mixing matrix that describes the  
linear combination of sterile and active neutrinos that participate in  
the solar neutrino oscillations.   
In the scheme of Eqs.~(\ref{eq:RCl}-\ref{eq:RSNOCC}), we are   
implicitly neglecting any small differences in the energy-dependent effects  
associated with the passage of active and sterile neutrinos through matter.  
We do not assign a  
normalization factor to the low-energy neutrinos because their flux  
uncertainty, which is constrained by the solar luminosity, is only  
1\% (see Table~\ref{tab:spectrum}). We also do not assign  
a normalization factor to the intermediate-energy neutrinos because it  
is likely that the uncertainties in this flux are well understood~\cite{poor}.  
  
\begin{figure}[t]  
\centering\leavevmode 
\mbox{\psfig{file=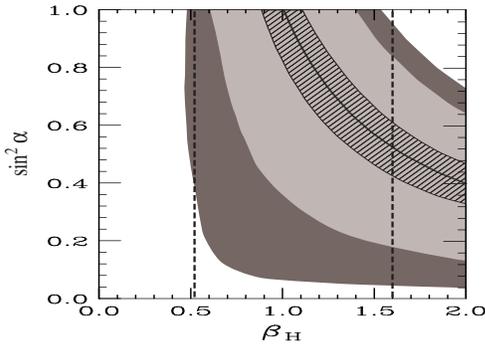,width=6.5cm,height=4.5cm}}  
\caption{Active neutrino fraction $\sin^2\alpha$  
versus $^8$B neutrino flux normalization $\beta_H$. The line  
represents solutions with $\chi^2=0$, and the $1\sigma$ and $2\sigma$   
allowed regions are shaded. The hatched area is the $1\sigma$  
allowed region if we replace the $R_{\rm SK}$ measurement by   
a hypothetical measurement $R^{\rm NC}_{\rm SNO}=1\pm 0.05$.   
The vertical dashes show the $3\sigma$ range allowed by the SSM.}  
\label{fig:vsbeta}  
\end{figure}

The solar neutrino data are summarized in Table~\ref{tab:data}.  
We note that before the recent SNO CC result, the $P_j$ were determined  
only if particular assumptions are made about the flux normalizations  
and sterile neutrino content~\cite{bmw}. With the addition of the SNO  
CC data, however, the quantities $\beta_H P_H$, $P_I$, and $P_L$  
can now be determined by $R_{\rm Cl}$, $R_{\rm Ga}$, and  
$R^{\rm CC}_{\rm SNO}$.  
We note that if the flux normalization $\beta_H$ were known,  
the $P_j$ would now be completely determined, regardless of the sterile  
content. This is because the $^{37}$Cl, $^{71}$Ga, and SNO CC  
measurements do not depend on whether the   
solar $\nu_e$ oscillate to active or sterile neutrinos.  
\begin{table}[t]  
\begin{eqnarray}  
\begin{array}{lc|cc}  
\rm{Experiment} & & &\rm{data/SSM}\\  
\hline  
^{37}\rm{Cl}  & & &0.337 \pm 0.030\\  
^{71}\rm{Ga} & & &0.584 \pm 0.039\\  
{\rm{Super\!-\!K}}  & & & 0.459 \pm 0.017 \\  
{\rm SNO~CC}  & & & 0.347 \pm 0.028 \nonumber  
\end{array}  
\end{eqnarray}  
\caption[]{\label{tab:data}  
Solar neutrino data expressed as the ratio $R =$~data/SSM, including the  
experimental uncertainties. The $^{71}$Ga number combines the results  
of the GALLEX, SAGE, and GNO experiments.}  
\end{table}  
The SK data may be used to further constrain the parameters  
$\beta_H$, $P_H$, and $\sin^2\alpha$, but without some assumption about  
either the $^8$B flux normalization or sterile neutrino content there  
will still be one unconstrained degree of freedom. Thus there exists a  
family of solutions that fit the data exactly (with $\chi^2 = 0$),  
described by the relation  
\begin{equation}  
\sin^2\alpha = {(R_{\rm SK} - R^{\rm CC}_{\rm SNO}) /  
\left[ r (\beta_H - R^{\rm CC}_{\rm SNO}) \right] } \,,  
\label{eq:sin2vsbeta}  
\end{equation}  
shown in Fig.~\ref{fig:vsbeta} by the solid curve. The amount of sterile  
content is not {\it a priori} known; in principle any value of  
$\sin^2\alpha$ between zero and unity is still possible.  
  
The $1\sigma$ and $2\sigma$ allowed regions from a fit to the rates in   
Table~\ref{tab:data} with five  
parameters ($\sin^2\alpha, \beta_H, P_H, P_I, P_L$) with the uncertainty in  
$\beta_H$ determined by the fit, are also shown in  
Fig.~\ref{fig:vsbeta}.  
A pure sterile oscillation solution ($\sin^2\alpha = 0$) is disfavored  
since experimentally \mbox{$R_{\rm SK} > R^{\rm CC}_{\rm SNO}$.}   
However, for large  
enough $\beta_H$, $\sin^2\alpha$ can be close to zero, although large flux  
normalizations are unlikely. For $\beta_H \le 2$   
(the 5$\sigma$ bound from the SSM), we find that the pure sterile case  
($\sin^2\alpha = 0$) is not acceptable at the $2\sigma$ level.  
At the 1$\sigma$ level and for $\beta_H \le 2$,   
the values obtained from the above analysis are   
\begin{eqnarray}  
\beta_H P_H &=& 0.35^{+0.07}_{-0.07} \,, \ P_I = 0.31^{+0.42}_{-0.31}  
\,,\ P_L = 0.81^{+0.19}_{-0.33} \,,\nonumber\\  
0.6 &\leq& \beta_H \leq 2\,, \ \ \ 0.14 \leq \sin^2\alpha \leq 1\,.    
\label{eq:three}  
\end{eqnarray}  
These can be used to make statements  
about particular models of neutrino oscillations.   
For example, the LMA  
solution has the ordering $P_H \le P_I \le P_L$, while for the Small  
Mixing Angle (SMA) solution $P_I$ is significantly suppressed below both  
$P_H$ and $P_L$. For  
$\beta_H \agt 1.14$ the probability hierarchy of the LMA solution can be  
satisfied. Furthermore, the measured SNO spectrum appears to be  
undistorted compared to the SSM, which favors the LMA solution. The LOW  
solution has $P_H = P_I = P_L$, and thus is disfavored. At  
$2\sigma$, the lowest allowed value of $\beta_H$ is 0.47,   
which occurs for pure  
active mixing ($\sin^2\alpha = 1$).  The vacuum solution with $\delta  
m^2 \sim 5.5 \times 10^{-12}\ {\rm{eV}}^2$ and large  
mixing~\cite{justso} is therefore barely acceptable at the $2\sigma$ level  
since the best-fit $\beta_H$ for this solution is $0.47$ (with very  
small uncertainties)~\cite{analysis}.  
  
In the near future SNO will also measure the NC reaction, which is related  
to the parameters by  
\begin{equation}  
R^{\rm NC}_{\rm SNO} = \beta_H P_H + \beta_H \sin^2\alpha (1 - P_H) \,.  
\label{eq:RSNONC}  
\end{equation}  
Equations~(\ref{eq:RSK}), (\ref{eq:RSNOCC}), and (\ref{eq:RSNONC})   
show that $R^{\rm NC}_{\rm SNO}$ does not provide independent  
information. There is in fact a universal sum rule:  
\begin{eqnarray}  
R^{\rm NC}_{\rm SNO} &=& \left[ R_{\rm SK} -  
(1 - r) R^{\rm CC}_{\rm SNO} \right]/r  
\nonumber \\  
&=& 5.85 R_{\rm SK} - 4.85 R^{\rm CC}_{\rm SNO} \,, 
\label{eq:sumrule}  
\end{eqnarray}  
that holds for any value of $\sin^2\alpha$ (this equation was   
known~\cite{bmw,analysis} for the case $\sin^2\alpha=1$). The SK and SNO data predict  
$R^{\rm NC}_{\rm SNO} = 1.00\pm 0.24$. Although the SNO NC  
measurement will not provide a new constraint because  
the SK data already supplies NC information,  
the SNO NC data could provide  
the more accurate measurement (since the $\nu_{\mu,\tau}$ NC  
cross sections are the same as that for $\nu_e$, unlike in SK  
where they are much less).   
If we replace $R_{\rm SK}$ by   
$R^{\rm NC}_{\rm SNO}=1\pm 0.05$ (in anticipation of a measurement accurate   
to 5-10\%~\cite{private}), we find $\sin^2\alpha>0.33$ at the  
1$\sigma$ level for $\beta_H \leq 2$.   
Another way to see why the SNO NC rate will not determine $\sin^2\alpha$   
is to consider the ratio  
\begin{equation}  
{R^{\rm NC}_{\rm SNO} / R^{\rm CC}_{\rm SNO}}=  
1+\sin^2\alpha\left( {1 / P_H}-1\right)\,.  
\end{equation}  
Since $P_H$ always appears in Eqs.~(\ref{eq:RCl})--(\ref{eq:RSNOCC}) in the combination $\beta_H P_H$,  
$\sin^2\alpha$ can not be extracted.

\begin{figure}[t]  
\mbox{\psfig{file=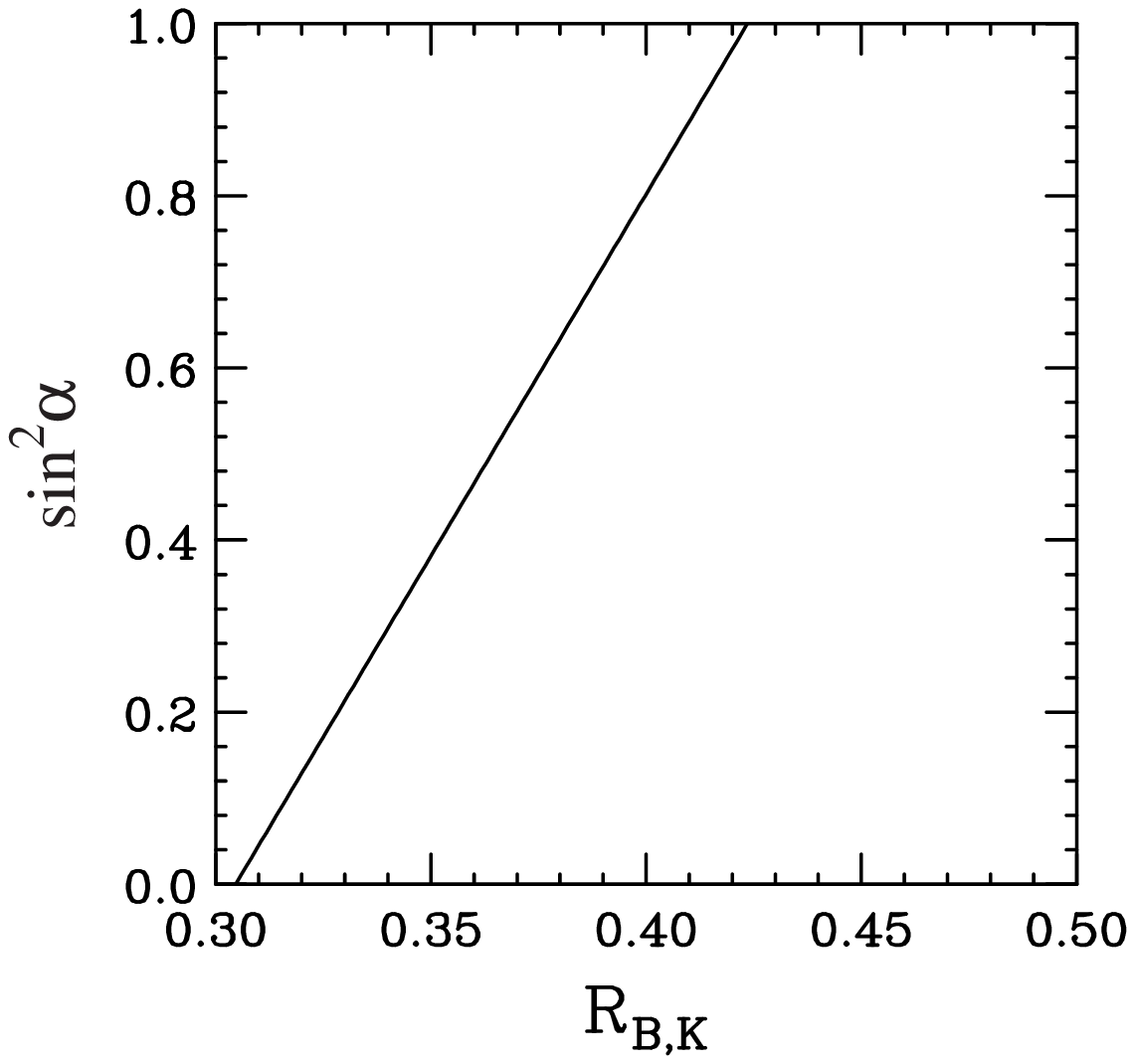,width=4.2cm,height=4cm}  
\psfig{file=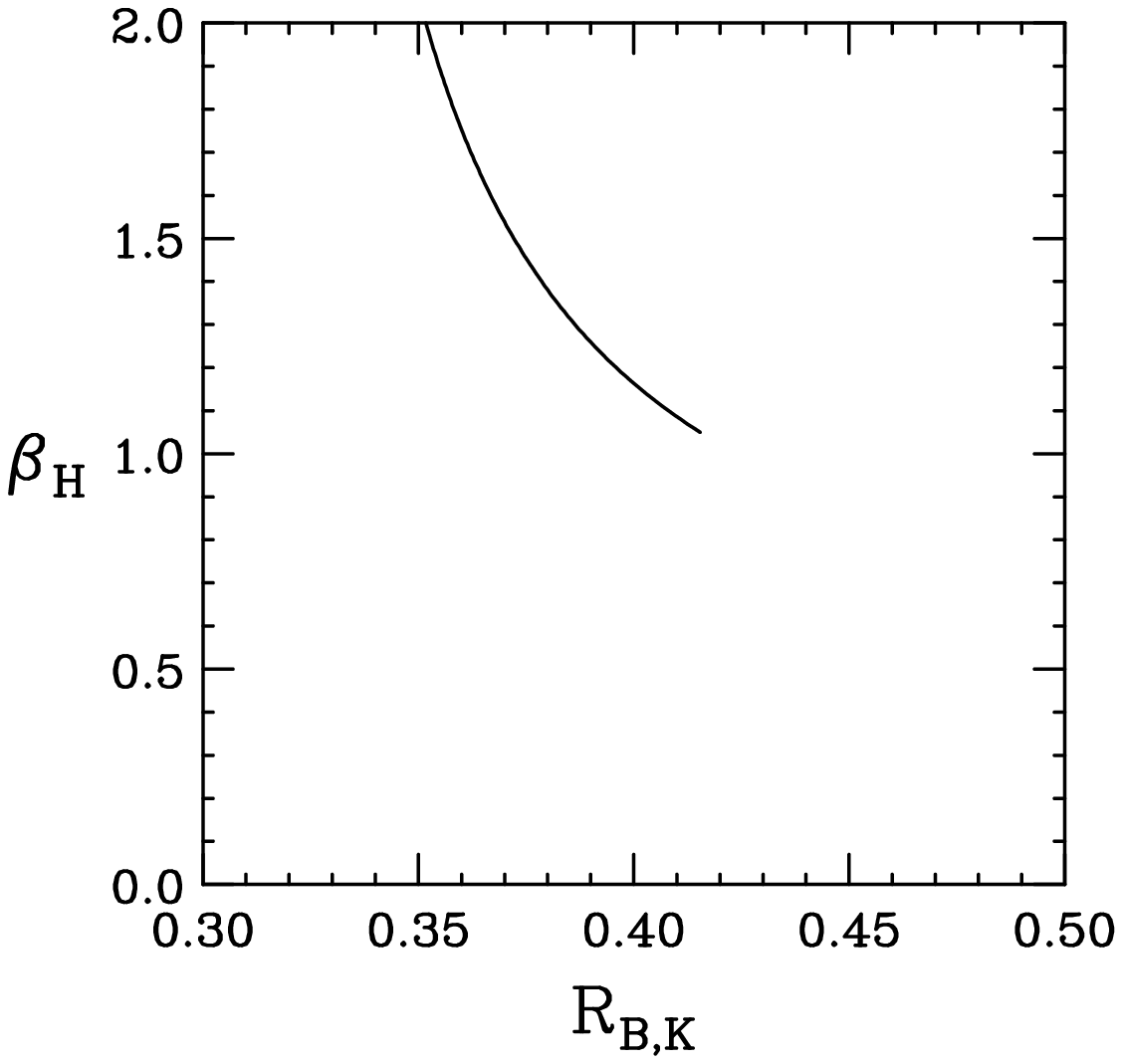,width=4.2cm,height=4cm}}   
\caption{Values of the active neutrino fraction $\sin^2\alpha$, and  
$^8$B neutrino flux normalization $\beta_H$ versus $R_{B,K}$.}  
\label{fig:vsrb}  
\end{figure}

\begin{figure}[t]  
\centering\leavevmode 
\mbox{\psfig{file=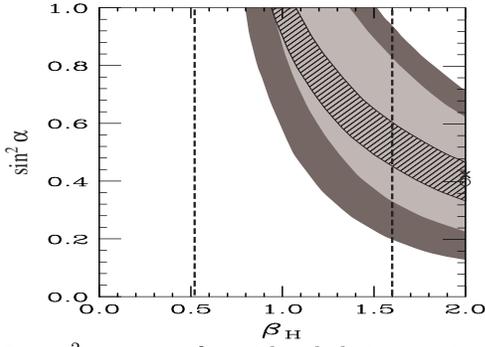,width=6.5cm,height=4.5cm}}  
\caption[]{$\sin^2\alpha$ versus $\beta_H$ with adiabatic   
constraints imposed. The cross marks the best-fit point   
$(\beta_H,\sin^2\alpha)$ $=(2.0,0.42)$ for $\beta_H \leq 2$;   
the $1\sigma$ and   
$2\sigma$ allowed regions are shaded. With $R_{\rm SK}$ replaced by  
$R^{\rm NC}_{\rm SNO}=1\pm 0.05$, the best-fit shifts slightly to the circle  
and the hatched region is the $1\sigma$  
allowed region. The dashed lines show   
the $3\sigma$ range allowed by the SSM.   
}  
\label{fig:adia}  
\end{figure}

\vskip 0.1in  
\noindent  
\underline{Can Borexino/KamLAND break the $\beta_H$, $\alpha$ degeneracy?}  
How then can the last degree of freedom ($\beta_H$ or $\sin^2\alpha$)   
be eliminated? To make a {\it model-independent} determination of   
both $\beta_H$ and  
$\sin^2\alpha$ (and hence also $P_H$), a different measurement that  
provides an  independent constraint on the parameters must be used. For  
example, a measurement of the intermediate-energy solar neutrinos that   
involves a NC contribution such as in the Borexino~\cite{borexino} experiment or in   
the solar neutrino component of the KamLAND~\cite{kamland} experiment,  
would allow a separate determination of $\sin^2\alpha$~\cite{murayama}.   
The resulting  
constraint would have the form  
%
$R_{B,K} = P_I + r \sin^2\alpha (1 - P_I)$,  
in analogy to Eq.~(\ref{eq:RSK}), and would determine $\sin^2\alpha$.  
Values of $\sin^2\alpha$ and $\beta_H$ (from Eqs.~(\ref{eq:RCl}),   
(\ref{eq:RSK}) and (\ref{eq:RSNOCC})) are shown versus $R_{B,K}$ in  
Fig.~\ref{fig:vsrb}. The value of $\beta_H$ does not extend below   
about $1$  
because $\sin^2\alpha$ becomes greater than unity there.  
It is difficult for Borexino or KamLAND to  
determine $\sin^2\alpha$ and $\beta_H$ because like SK, there is  
limited sensitivity to the NC component of the detected flux;   
the resulting uncertainty in $\sin^2\alpha$ would be $\delta  
\sin^2\alpha = \delta R_{B,K}/[r(1-P_I)] \simeq 8 \delta R_{B,K}$.  
\vskip 0.1in  
\noindent  
\underline{Can adiabatic constraints break the $\beta_H$, $\alpha$ degeneracy?}  
If a particular model is assumed, then it can provide the additional  
constraint to determine the parameters from current data. For example,  
if in the LMA solution all of the high energy   
neutrinos and a  
fraction $f$ of the intermediate energy neutrinos are created above  
resonance, and a fraction $1-f$ of the intermediate energy neutrinos and  
all of the low energy neutrinos are created below resonance, then since  
the neutrinos propagate adiabatically ({\it i.e.} the probability of  
jumping across the Landau-Zener-type level crossing   
from one adiabatic state to another is small) in the Sun we have  
approximately  
\begin{eqnarray}  
P_H &=& \sin^2\theta \,,\ \ \ \ \ \ \ \  P_L = 1- \textstyle {1\over2}\sin^22\theta \,,   
\label{eq:PHLMA}\\   
P_I &=& f \sin^2\theta + (1-f)(1- \textstyle {1\over2}\sin^22\theta) \,,  
\label{eq:PILMA}  
\end{eqnarray}  
where $\theta$ is the vacuum mixing angle and  
$f$ can be directly related to the solar $\delta m^2$ (for a more detailed  
discussion, see Ref.~\cite{bmw}). Note that  
Eqs.~(\ref{eq:PHLMA}) and (\ref{eq:PILMA}) imply $P_H \le P_I \le  
P_L$ (for $\sin^2\theta \le 1/2$). Now there are only  
 four parameters ($\beta_H$,  
$\sin^2\alpha$, $f$, and $\theta$) and all can be determined from
the present data.  
Constraining $\beta_H \leq 2$, we find the best-fit point to be  
%
\begin{eqnarray}  
\beta_H &=& 2.0 \,, \ \sin^2\alpha = 0.42 \,,\  \sin^2\theta=0.17\,,  
\ f = 0.6\,, 
\label{eq:LMA}   
\end{eqnarray}  
with $\chi^2=0.51$. (There is a unique solution with zero $\chi^2$, but   
has $\beta_H=3.26$, which is unreasonably high~\cite{high}).   
The 1$\sigma$ and 2$\sigma$ allowed regions from a  
four-parameter fit are shown in Fig.~\ref{fig:adia}.   
Note the similarity  
of the regions of Figs.~\ref{fig:vsbeta} and~\ref{fig:adia};   
adiabatic constraints do not greatly help reduce the allowed region.   
  
\begin{figure}[t]  
\mbox{\psfig{file=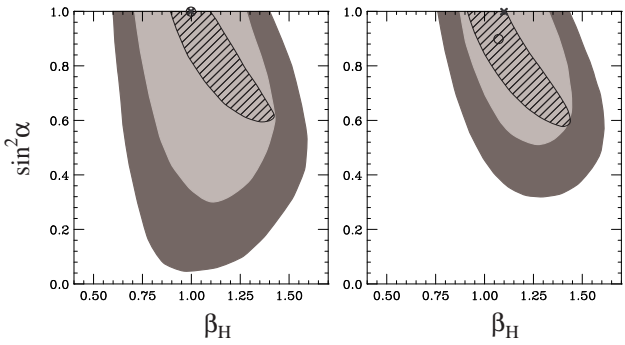,width=8.5cm,height=4.5cm}}  
\caption[]{$\sin^2\alpha$ versus $\beta_H$ with the $^8$B flux    
constraint imposed. The left panel shows the result of a model-independent  
analysis, and the right panel has adiabatic constraints in addition.   
The crosses mark the best-fit points;   
the $1\sigma$ and   
$2\sigma$ allowed regions are shaded. The hatched areas are the $1\sigma$  
allowed regions if we replace $R_{\rm SK}$ by  
$R^{\rm NC}_{\rm SNO}=1\pm 0.05$; the circles mark the corresponding  
best-fit points.}  
\label{fig:nadia}  
\end{figure}

\vskip 0.1in  
\noindent  
\mbox{\underline{Including the SSM constraint on the $^8$B flux.}  
To include} the $^8$B flux normalization as calculated in the SSM~\cite{SSM},   
we perform $\chi^2$ analyses with $\beta_H=1\pm 0.18$.  
The results are shown in Fig.~\ref{fig:nadia}.    
The model-independent analysis yields a unique point with $\chi^2=0$ at   
$(\beta_H,\sin^2\alpha)$ $=(1.0,1.0)$. The   
left panel of Fig.~\ref{fig:nadia} shows that     
the $\sin^2\alpha$ range is not improved. However, as shown in the right panel   
of Fig.~\ref{fig:nadia}, imposition of the $^8$B flux constraint  
in addition to adiabatic constraints, does lead to a smaller  
$\sin^2\alpha$ range.  
In this case, the best-fit parameters are  
\begin{eqnarray}  
\beta_H &=& 1.1 \,, \ \sin^2\alpha = 1.0 \,,\  \sin^2 2\theta=0.83\,,  
\ f = 0.15\,, 
\label{eq:consLMA}   
\end{eqnarray}  
with $\chi^2=3.1$. Thus,   
for this solution mainly high energy  
neutrinos are created above resonance and the critical energy~\cite{bmw}   
 lies close to the $pep$ line at 1.44 MeV,  
which translates to \mbox{$\delta m^2 = 4.8 \times 10^{-5}\ {\rm{eV}}^2$.}  
%
%
\vskip 0.1in  
\noindent  
\underline{Summary.}  
After including the recent SNO CC results in a model independent  
analysis of solar neutrino flux rate data, there remains one free  
parameter. The locus of solutions may be represented by a curve in the   
 plane of the  
the active neutrino fraction, $\sin^2\alpha$, and the $^8$B neutrino flux  
normalization $\beta_H$. We have shown that the forthcoming SNO NC data will  
not fully  constrain the last degree of freedom; in fact, there is a
universal sum rule  
involving $R^{NC}_{\rm SNO}$, $R^{CC}_{\rm SNO}$, and $R_{\rm SK}$ that  
must be satisfied, independent of the sterile neutrino content of the  
solar neutrino flux. The adiabatic constraint for the LMA  
does not appreciably reduce the allowed region in $\sin^2\alpha$.   
Even when we impose the SSM $^8$B flux constraint, the sterile  
neutrino fraction is not determined.  
  
In principle, measurements of $\nu_e$ scattering for the  
intermediate-energy neutrinos in Borexino/KamLAND could break the  
degeneracy of allowed solutions, but because the NC sensitivity of these  
experiments is relatively weak, a very precise measurement would be  
required to determine $\sin^2\alpha$ and $\beta_H$. What is needed is a  
measurement of neutrino-nucleon NC scattering for the  
intermediate-energy neutrinos.  
  
{\it Acknowledgments}:  
This research was supported by the U.S.~DOE  
under Grants No.~DE-FG02-95ER40896 and No.~DE-FG02-01ER41155   
and by the WARF.

\end{document}